\documentstyle[12pt]{article}
\newcommand{\calA}{${\cal A}$}

\begin{document}

\title{\Large \bf Origin of Classical Singularities}

\author{Michael 
Heller\\Vatican Observatory, V-00120 Vatican City State\\ 
Correspondence address:\\  ul.  Powsta\'nc\'ow Warszawy 13/94 \\ 33-110 
Tarn\'ow, Poland\\
e-mail mheller@wsd.tarnow.pl \\
\and
Wies{\l}aw Sasin\\ Technical University of 
Warsaw, \\
Plac Politechniki 1, 00-661 Warsaw, Poland}

\maketitle

\begin{abstract}
We briefly review some results concerning the problem of
classical singularities in general relativity, obtained with the
help of the theory of differential spaces. In this theory one
studies a given space in terms of functional algebras defined on
it. Then we present a generalization of this method consisting
in changing from functional (commutative) algebras to
noncommutative algebras. By representing such an algebra as a
space of operators on a Hilbert space we study the existence and
properties of various kinds of singular space-times. The
obtained results suggest that in the noncommutative regime,
supposedly reigning in the pre-Planck era, there is no
distinction between singular and non-singular states of the
universe, and that classical singularities are produced in the
transition process from the noncommutative geometry to the
standard space-time physics.
\end{abstract}
\section{Introduction}
\par The standard method of dealing with singularities in
general relativity consists in probing the incompleteness of
space-time (geodesic incompleteness, Schmidt's b-incompleteness
or some other suitably chosen incompleteness) with the help of
various structures defined on the space-time manifold (such as
the chronological or causal structures) and certain conditions
superimposed on it (such as different energy conditions). In
this approach singularities are treated as ideal or boundary
points of space-time which can be reached only by investigating
regular (i.e., nonsingular) regions of space-time (see
\cite{HawEl,SingRev}). In the series of works
\cite{GH92,GH93,HS91,HS94,HS95JMP,HS95AC,HSTZ} we have proposed
an alternative method of investigating singularities in general
relativity. The idea goes back to Geroch \cite{Geroch} who
suggested that the space-time geometry, instead of being studied
in terms of the usual charts and atlases on a manifold $ M$, can
be investigated in terms of the algebra $C^{\infty} (M)$ of
smooth functions on $M$. Although this method is, in principle,
equivalent to the traditional one, it is more algebraic and more
global as far as its intuitive character is concerned and, first
of all, it seems to be more open for further generalizations.
The tempting idea is to consider a more general functional
algebra $\bar {C}$ and treat it as {\em - ex definitione -\/} an
algebra of smooth functions on a larger space $
\bar {M}$ such that if we restrict $\bar {C}$ to $M$
we obtain the standard geometry on a smooth manifold $M$. One
could hope that the set $\bar { M}\setminus M$ contains
space-time singularities which, in this case, would be directly
accessible to investigations in terms of the algebra $\bar {
C}$.
\par
It turns out that such a program can indeed be implemented
provided that the algebra $\bar {C}$ is subject to some further
conditions which would allow one to define standard differential
tools in terms of it.  The pair $(\bar {M},\bar {C})$, where
$\bar C$ is a functional algebra satisfying suitable conditions,
is usually called {\em differential space}. The theory of these
spaces is a quickly developing field of research in mathematics
(for a bibliographical review see \cite{Buchner}; in \cite{H92},
to develop the above sketched program we have used the theory of
differential spaces elaborated by Sikorski
\cite{Sik67,Sik71,Sik72}).
\par
In Section 2, to make the paper self-contained, we briefly
review the main results concerning the problem of singularities
obtained with the help of the theory of differential spaces and,
in Section 3, we explicitly show a geometrical mechanism of the
formation of singularities, and illustrate it with some
examples. In this respect, the functional algebra method seems
to be more efficient than the traditional one. However, also
this method trivializes if we try to apply it to the
investigation of stronger type singularities.  It turns out that
the functional algebra method can be generalized by replacing
commutative algebras by noncommutative ones. In Section 4, we
effectively construct such a noncommutative algebra \calA , and
in Sections 5 and 6, we show how can it be used to the study of
singularities. Even the strongest singularities (which we call
malicious singularities) surrender to this method. We study the
existence and structure of malicious elementary quasiregular and
regular singularities.  The obtained results can be interpreted
in the following way.  The pre-Planck era is modelled by a
noncommutative geometry, and in this era there is no distinction
between singular and nonsingular states of the universe. Passing
through the Planck threshold consists in changing from the
noncommutative algebra to (a subset of) its center. In this
process the standard space-time physics emerges and classical
singularities are produced. Consequently, the question of the
existence of singularities at the fundamental level is
meaningless.  It is only from the point of view of the
macroscopic observer that one can ask whether the universe had
the initial singularity in its finite past or will have the
final singularity in its finite future.
\par
\section{Singularities in Terms of Fun\-ctional Algebras}
Let us consider a family $C$ of real valued functions on a set $
M$ which we endow with the weakest topology $\tau_C$ in which
the functions of $ C$ are continuous.
\par
A function $f$, defined on $A\subset M,$ is called a {\em local
$C$-function\/} if, for any $x\in A, $ there is a neighborhood $
B$ of $x$ in the topological space $(A, \tau_A)$, with $\tau_ A$
the topology induced in $A$ by $\tau_C$, and a function $g\in C$
such that $g|B=f|B$. Let $ C_A$ denote the set of all local
$C$-functions. It can be easily seen that $C\subset C_M$; if
$C=C_M$, the family $C$ is said to be {\em closed with respect
to localization}.
\par
$C$ is said to be {\em closed with respect to superposition with
smooth Euclidean functions\/} if for any $n\in {\bf N}$ and each
function $\omega\in C^{\infty}({\bf R}^n),\;f_1,...,f_ n\in C$
implies $\omega\circ (f_1,...,f_n)\in C$. It is easy to see that
this condition implies that $C$ is a linear algebra.
\par
A family $C$ of real valued functions on $M$ which is both
closed with respect to localization and closed with respect to
superposition with Euclidean functions is called a {\em
differential structure\/} on $ M$, and a pair $(C,M)$, where $C$
is a differential structure on $ M$, is called a {\em
differential space}. Of course, every differential manifold is a
differential space with $C=C^{\infty}(M)$ as its differential
structure.
\par
The above construction is more general and more flexible if we
use a sheaf ${\cal C}$ of linear functional algebras on a
topological space $ M$ (with any topology $\tau$) instead of a
single functional algebra $C$ on $ M$ (with the $\tau_C$
topology).  In such a case, the condition of the closeness with
respect to localization is already contained in the sheaf
axioms. The triple $ (M,\tau ,{\cal C})$ is called a {\em
structured space}, and ${\cal C}$ its {\em differential
structure}. The theory of these spaces was developed in
\cite{HS94,HS95JMP}. 
\par
It is evident that any differential space can be trivially
regarded as a structured space, but not vice versa. It can be
shown that a structured space $(M,\tau ,{\cal C})$ is a
differential space if for every open set $U\in\tau$ and any
point $x\in U$, there exists a function $
\varphi$, called {\em bump function}, such that $\varphi (p)=1$
and $\varphi |M\setminus U=0.$
\par
A space-time with singularities (i.e., with a singular boundary)
can be organized into a structured space and, in this way,
singularities can be investigated with the help of the theory of
structured spaces.  This is done by using the method of
prolongations of differential structures. Let $ M$ be a
space-time manifold, which is, of course, also a structured
space $ (M, \tau ,{\cal C})$ such that $\tau$$=\tau_{{\cal C}}$
where $\tau_{{\cal C}}$ is the weakest topology in which
functions of $ {\cal C}$ are continuous. Let $\bar
{M}=M \cup \partial M$ where $\partial M$ is a singular
boundary of $M$. A sheaf $\bar {{\cal C}}$ on $\bar {M}$ such
that $
\bar {{\cal C}}(M)={\cal C}(M)$ is said to be a 
{\em prolongation\/} of the differential structure $ {\cal C}$
on $M$ to that of $\bar {M}.$ Since $M$ is dense in $\bar {M}$
the prolongation is unique. It turns out that such a
prolongation always exists although in 
some cases it is trivial (see below).
\par
In the study of singularities we assume that $M$ is a space-time
and $\bar M$ this space-time with singularities.  In this way,
regular singularities and elementary quasiregular singularities
(in the classification of Ellis and Schmidt \cite{EllisSch}) can
be fully analyzed \cite{HS91,Odrzy}). In particular, the conic
singularity in the space-time of a cosmic string has been
thoroughly studied by using this method \cite{GH92,HS95AC}, and
the results have been found consistent with those obtained by
Vickers \cite{Vicka,Vickb,Vickc} who used other methods. These
are the easiest cases to deal with. As an another extremity, the
situations have been studied in which the singular boundary is
not Hausdorff separated from the rest of space-time. It has been
demonstrated that in such cases only constant functions can be
prolonged from the differential structure of a given space-time
to its singular boundary \cite{HS94,HS95JMP}. As it is well
known, situations of this kind occur in the closed Friedman
world model and in the Schwarzschild solution when their
curvature singularities are interpreted as Schmidt's
b-boundaries. In the closed Friedman world model another
pathology occurs: both the initial and final singularities turn
out to be the same and the only point of the b-boundary of this
space-time \cite{Bosshard,Johnson}. This situation is
transparently explained in terms of structured spaces: the
differential structure ${\bar C}({\bar M})$ of the closed
Friedman space-time $M$ together with its b-boundary $
\partial M$, $\bar {M}=M\cup\partial M$, consists only of
constant functions, $\bar {{\cal C}}(\bar {M})\simeq {\bf R}$,
which do not distinguish between points, i. e., there is no
function $ f\in\bar {{\cal C}}(\bar {M})$ such that $f(p)\neq
f(y)$ for any $x,y\in\bar { M},$$\,x\neq y$. However, if we do
not ``touch'' any singularity, i. e., if we consider the
differential structure ${\cal C}(M)$ of space-time $M$ rather
than $\bar { {\cal C}}(\bar {M})$ everything remains all right
\cite{HS94,HS95JMP}. 
\par
More generally, we have proved that if $x_0$ is a b-boundary
point, and if the fiber $\pi^{-1}(x_0)$ in the (generalized)
fiber bundle of linear frames over $\bar {M}=M\cup\partial_bM$,
$x_0 \in \partial_bM$, degenerates to a single point, then the
only global cross sections of the sheaf $\bar {{\cal C}}$ over
$\bar {M}$ are constant functions, i. e., $\bar {{\cal C}}(\bar
{M})\simeq {\bf R}$ (see \cite{HS95JMP}). In such a case $x_0$
is called {\em malicious singularity.\/} Singularities of the
closed Friedman universe and of the Schwarzschild solution
belong to this class of singularities.
\par
\section{Origin of Classical Singularities in Terms of 
Differential Spaces} 
In this section we shall study a ``mathematical mechanism'' of
the appearance of classical singularities, and illustrate it
with some examples.
\par
Let $(M,C)$ be a differential space (in the sense of Sikorski),
and  $\rho\subset M\times M$ an equivalence relation in $M$. The
family
\[\bar {C}:=C/\rho =\{\bar {f}:\,M/\rho\rightarrow 
{\bf R}:\bar {f}\circ\bar{\pi}_{\rho}\in C\},\] where
$\pi_{\rho}:M\rightarrow M/\rho :=\bar { M}$ is the canonical
projection, is the largest differential structure on $M/\rho$
such that $\pi_{\rho}$ is smooth (in the sense of the theory of
differential spaces). When going from the set $M$ to the set
$\bar {M}$ some elements of $ M$ are glued together forming
various kinds of singularities. We shall say that a function
{\em passes through\/} a singularity if the singularity is in
its domain. It is obvious that the differential structure $\bar
{ C}$ is the maximal set of functions passing through
singularities in the quotient space $\bar {M}=M/\rho$. Let us
define another linear algebra of functions
\[C_{\rho}:=\{f\in C:\forall_{x,y\in M}x\rho 
y\Rightarrow f(x)=f(y)\}.\]
Any function belonging to $C_{\rho}$ is called $
\rho$-{\em consistent}. There exists an isomorphism of linear
algebras $\Phi :\,\bar {C}\rightarrow C_{\rho}$ given by $\Phi
(\bar {f})=\bar {f}\circ\pi_{\rho}$ for $\bar {f}\in\bar {C}$
(all these facts are proven in \cite{HSTZ}). As we can see,
there is an isomorphism between functions passing through
singularities and $\rho$-consistent functions.
\par
In many cases considered in general relativity, the equivalence
relation $\rho $ is defined by the action of a group $\Gamma$ on
$ M$: $M\times\Gamma\rightarrow M$. In such a case, for $x,y\in
M$,
$$x\rho y\Leftrightarrow \exists g\in\Gamma\;{\rm such}\; {\rm
that}\; y=xg.$$ 
The family $C_{\Gamma}=\{f\in C:\forall x\in
M,\,g\in\Gamma ,f(x)=f(xg)\}$ of functions belonging to $C$,
which are constant on the orbits of $ \Gamma$, is called the
family of $\Gamma${\em -invariant functions}. It is clearly a
linear algebra.
\par
To summarize the above analysis we can say that singularities
are formed from the initially smooth space-time modelled by a
differential space $ (M,C)$, where $M$ is a smooth manifold and
$C=C^{\infty}( M)$, in the process of forming the quotient space
$\bar {M}=M/\rho$. The space-time with singularities is now
modelled by the quotient differential space $(M/\rho ,C/\rho )$
such that $C/\rho $ is isomorphic with $C_{\rho}$. This schema
can be visualized in the form of the following diagram
\[(M,C)\Rightarrow (M/\rho ,C/\rho )\Leftrightarrow 
(M,C_{\rho}).\]
\par
We shall illustrate this schema with some simple examples.  
\par 
{\bf Example 1:  Regular singularity.}  Let $ M={\bf R}^2$ and
$\Gamma ={\rm O}(2)$.  The differential space without
singularities is $ ({\bf R}^2,C^{\infty}({\bf R}^2)${\bf .} Let
us define the equivalence relation $\rho $ in the following way:
$p\rho q$, $p,q \in M$, iff there exists $g \in \Gamma $ such
that $q = pg$. When we go to the quotient differential space
$(M/\rho ,$$C /\rho )$ singularities are formed.  Equivalence
classes of $\rho$ have the form of the concentric circles with
the degenerate circle, the singularity, at the origin.  The
singularity is a fixed point of the action of $\Gamma$, i.  e.,
the isotropy subgroup $\Gamma_{(0,0)}$ of the point $(0,0)$
coincides with the entire group
$\Gamma$.  The isotropy group of other points of $ M/\rho$ is
trivial, i.  e.,  $\Gamma_{(x,y)}=\{{\bf I}\}$ for all $
(x,y)\neq (0,0)$.  Now, we define the linear algebra of
$\Gamma$-invariant functions
\[C_{\Gamma}=\{f\in C^{\infty}({\bf R}^2):f(x
,y)=\omega (x^2+y^2),\,(x,y)\in {\bf R}^2,\,\omega
\in C^{\infty}({\bf R}^2)\}.\]
The isomorphism $\Phi :C_{}/\rho\rightarrow$$ C_{\Gamma}$ is
given by
\[\Phi [(x,y)]= x^2+y^2.\]
The points along the concentric circles are now identified and
the quotient space becomes a half-line ${\bf R}_{+}
.$ We obtain the quotient 
differential space $({\bf R}_{+},C^{\infty}({\bf R}_{ +}))$
which originates from cutting off parts of the original
differential space $({\bf R}^ 2,C^{\infty}({\bf R}^2))$, and
consequently the singularities which are formed in this
construction are regular singularities.
\par
{\bf Example 2:  Quasiregular singularity.}  Let $ M={\bf R}^2$
and $\Gamma =\{{\rm O}_{\frac {2\pi}3},{\rm O}_{\frac
{4\pi}3},{\rm O}_0)$, the rotation group by the angles $2\pi
/3,\,4\pi /3,\,0$, respectively. The set of fixed points of the
action of $
\Gamma$ consists of one point $(0,0)$, and each orbit, for
nonsingular points, consists of three points. The isotropy
subgroups are
\[\Gamma_{(x,y)}=\left\{
\begin{array}{ll}
\Gamma & \mbox{for $(x,y)=(0,0)$} \\
{\bf I} & \mbox{for $(x,y)\neq (0,0)$}
\end{array}
\right.
\]
After making suitable identifications we obtain a cone. For
details see \cite{HS95AC}.
\par 
{\bf Example 3:  Malicious singularity.}  We recall that a
singularity is called malicious if the fiber over it in the
generalized frame bundle over the space-time with such a
singularity degenerates to a single point.  Such singularities
occur in the closed Friedman world model and in the
Schwarzschild solution.  In \cite{HS95JMP} we have
shown that the only functions that pass through the singularity
are constant functions, i.  e.,  in such a case $ C_{\rho}\simeq
{\bf R}$.  Since constant functions do not distinguish points,
the differential space $ (M/\rho ,C/\rho )$ modelling space-time
with a malicious singularity is diffeomorphic with the space
$(\{{\rm p}{\rm o}{\rm i}{\rm n}{\rm t}
\},\,${\bf R}). This also explains why the b-boundary of a
space-time with malicious singularities consists of a single
point. For instance, such a situation occurs for the closed
Friedman model in which the initial and final singularities are
the same and the unique point of the b-boundary
\cite{Bosshard,Johnson}. For this model $C_{\rho } \simeq {\bf
R}$, and consequently only zero vector fields can be prolonged
to the b-completion of the corresponding space-time. Therefore,
any curve joining the initial and final singularities must have
the zero ``bundle length''. (For details see
\cite{HS94,HS95JMP}.)
\par
The analysis carried out in the present section can also be done
in terms of structured spaces. The method of using sheaves of
functional algebras rather than algebras is even more flexible
since it does not presuppose {\em a priori\/} fixed topology.
Nevertheless, the main difficulty remains:  although the source
of difficulties with malicious singularities is beautifully
explained, their structure is not analyzable with the help of
this method.
\par 
\section{Desingularization Procedure} 
To gain more insight into what happens in malicious
singularities we have proposed to replace commutative functional
algebras, regarded as differential structures, by noncommutative
ones \cite{HS96,Banach}. The idea is to generalize a space-time
$ M$ with a singular boundary $\partial M$, $\bar
{M}=M\cup\partial M$, to a noncommutative space in the sense of
Connes \cite{Connes}. Such a space is essentially nonlocal, and
when the above generalization is done we loose information on
single points, but we gain the information about states, and
both ``singular'' and ``nonsingular'' states are on equal
footing. In this sense, the proposed construction can be
regarded as a desingularization procedure. In doing it we
closely follow the method described by Connes
\cite[p.99]{Connes} and elaborated by him for other purposes.
\par
Let $M$ be a space-time and $\overline {OM}$ the Cauchy
completed total space of the orthonormal frame bundle over M
with the structural group $\Gamma ={\rm S}{\rm O}(3,1)$. Then
$\bar {M} =\overline {OM}/\Gamma$ is the b-completion of the
space-time $ M$, and $\partial_bM=\bar {M}\setminus M$ is its
b-boundary. The Cartesian product $ G=\bar {M}\times\Gamma$ has
the structure of a groupoid [in this case, it can be called
a {\em (generalized) transformation groupoid\/}]. The elements
of $ G$ are pairs $\gamma =(p,g)$ where $p\in\overline {OM}$ and
$ g\in\Gamma ,$ and evidently two such pairs
$\gamma_1=(p_1,g_1)$ and $\gamma_2=(p_2,g_2)$ can be composed if
$ p_2=p_1g_1.$ If we represent $\gamma =(p,g)$ as an arrow
beginning at $ p$ and ending at $pg$, then two arrows $\gamma_1$
and $\gamma_2$ can be composed if the beginning of $
\gamma_2$ coincides with the end of $\gamma_1.$ Let us notice
that the ``frame'' $ p_0$ belonging to the ``singular fibre'',
i. e., $ p_0\in\pi^{-1}(x_0)$ where $x_0\in\partial_bM$, is not
an ordinary frame but rather the limit of Cauchy sequences of
orthonormal frames (see \cite{HS96}). From Schmidt's
construction it follows that such limits always exist
\cite{Schmidt}.
\par
Let us define two functions: beg$(\gamma )=p$ and end$
(\gamma )=pg$ for $\gamma =(p,g)\in G$. It is immediate to see
that the sets of all arrows that begin at $p\in\overline {OM}$
\[G^p:=\{\gamma\in G:\,{\rm b}{\rm e}{\rm g}(
\gamma )=p\}=\{(p,g):g\in\Gamma \}\]
and the set of all arrows that end at $q$$\in\overline {OM}$
\[G_p:=\{\gamma\in G:\;\,{\rm e}{\rm n}{\rm d}
(\gamma )=q\}=\{(qg^{-1},q):g\in\Gamma \}\]
can be given the structure of the group manifold $
\Gamma =$SO(3,1) even if $p$ or $q$ belong to the fiber over a
malicious singularity. In this way, also malicious singularities
can be represented by well behaved structures. This is another
reason for the name ``desingularization'' procedure.
\par
We are now ready to define the noncommutative involutive algebra
${\cal A} = (C_c^{\infty}(G,{\bf C}), *, ^{*})$ of compactly
supported complex valued functions on the groupoid $G$. The
multiplication in this algebra is defined to be the convolution
\[(s*t)(\gamma )=\int_{G_p}s(\gamma_1)t(\gamma_
1^{-1}\gamma )d\gamma_1\]
for every $s,t\in {\cal A}$, where $\gamma =\gamma_
1\circ\gamma_2$, $G_p$ is the fiber over $p\in\overline { OM}$,
and the integral is taken over the Haar measure on $G_p$. The
involution is defined in the following way
\[s^{*}(\gamma )=\overline {s(\gamma^{-1})}\]
where $\gamma^{-1}$ is the ``reversed arrow''. Geometry based on
the algebra $ {\cal A}$ should be regarded as a noncommutative
version of the space-time geometry with singularities. Indeed,
when ${\cal A}$ is restricted to the nondegenerate part of $G$
(to the nondegenerate fibres of $ G$) it is strongly Morita
equivalent to the algebra $C^{\infty}(M)$ of smooth functions on
the space-time manifold $M$ (strong Morita equivalence can be
regarded as a noncommutative counterpart of isomorphism, see
\cite[p.179]{Landi}). The states of the algebra ${\cal A}$, i.
e., positive and suitably normed linear functionals on ${\cal
A}$, represent states in the physical sense.  The algebra ${\cal
A}$ is `desingularized'', i. e., it does not distinguish between
``singular'' and ``nonsingular states'' \cite{Banach}.
\par
It is the standard thing to represent a noncommutative algebra
in a Hilbert space ${\cal H}$ and to study it in terms of the
algebra of bounded operators on ${\cal H}$. Following Connes
\cite[p.102]{Connes}, we define, for every $q \in
\overline{OM}$, the following representation of the algebra
${\cal A}$ 
\begin{equation}\label{repr}\pi_q:\,{\cal A}\rightarrow 
{\cal B}({\cal H})\end{equation}
by
\[(\pi_q(s)(\xi )=(s_q*\xi ),\]
for every $s\in {\cal A}$, where $\xi\in L^2( G_q):={\cal H}$.
This representation is involutive and nondegenerate. The
completion of ${\cal A}$ with respect to the norm
\[\parallel s\parallel ={\rm s}{\rm u}{\rm p}_{
q\in\overline {OM}}\parallel\pi_q(s)\parallel\]
is a $C^{*}$-algebra. As we shall see in the following,
singularities (even malicious ones) can be studied in terms of
these representations.
\par
To make contact with the analysis carried out in the preceding
Section in terms of commutative differential structures on
space-time it is important to answer the question when the
commutative algebra ${\cal A}$ reproduces the commutative
differential structure on a manifold.  Let us consider two
fibres $G_p$ and $G_q$, $ p,q\in
\overline{OM},$ of the groupoid $G$.  They are said to be {\em
equivalent\/} if there is $ g\in\Gamma$ such that $q=pg$.  A
function of ${\cal A}$ which is constant on the equivalence
classes of equivalent fibres is said to be {\em projectible.\/}
The set of all such functions, denoted by ${\cal A}_{proj}$,
forms a subalgebra of $ {\cal A}$.  It can be easily seen that
if $f,g\in {\cal A}_{proj}$, then their convolution $ f*g$
becomes the usual (commutative) multiplication, and ${\cal A}_{
proj}\subset {\cal Z}({\cal A})$ where ${\cal Z} ({\cal A})$ is
the center of ${\cal A}.$ One can readily show that $ {\cal
A}_{proj}$ is isomorphic to an algebra of complex valued
functions on $\bar { M}$.  It is evident that if there are no
singularities then ${\cal A}_{proj}$ is isomorphic with $
C^{\infty}(M)$.  If ${\cal A}_H$ denotes all Hermitian elements
of ${\cal A}$, then $ {\cal A}_{proj}\cap {\cal A}_H$ is
isomorphic with a family of real valued functions on M.
\par
\section{Nonlocal Geometry of Space-Time with Singularities} 
As it is well known (see, for instance, \cite[p.11]{Landi}),
every commutative $C^{*}$-algebra $C$ corresponds to a Hausdorff
topological space $M$ in the sense that $C$ is isometrically $
*$-isomorphic to the algebra of (complex valued) functions on
$M$.  Indeed, let us consider the space $\hat {C}$ of {\em
characters\/} of the algebra $ C$, i.  e.  the space of
functionals $\phi :\,C\rightarrow {\bf C}$ such that $
\phi (fh)=\phi (f)\phi (h)$, and equip it with the 
topology of pointwise convergence.  If $f\in C,$ then the
mapping $\hat {f}:\,\hat {C}\rightarrow {\bf C}$ defined by
$\hat {f}(\phi )=\phi (f)$, for all $
\phi\in\hat {C}$, is a continuous complex valued 
function on $\hat {C}$ (called the {\em Gel'fand
transform\/} of $f$).  The Gel'fand-Neimark theorem states that
all continuous functions on $ \hat {C}$ are of this form (for
some $f\in C$).  Consequently, one can regard elements of $C$ as
complex valued functions on $ \hat {C}$.  Equivalently, one can
define $\hat {C}$ as the set of maximal ideals of the algebra $
C$ with the Jacobson topology \cite[p.12]{Landi}. Every such
ideal is the set of functions vanishing at some point $x\in M$.
\par
The above construction does not work for noncommutative
algebras.  In general, such algebras have no maximal ideals, and
the structure which is the closest one to the concept of point
is the {\em primitive ideal\/} of a noncommutative algebra
${\cal A}$, i. e.,  the kernel of an irreducible
$*$-representation of ${\cal A}$.  If $ {\cal A}$ is a
$C^{*}$-algebra, $\pi$ a representation of ${\cal A}$ in a
Hilbert space $ {\cal H},$ and $\xi\in {\cal H}$, then $f\mapsto
(\pi (f)\xi ,\xi )$ is a positive form on ${\cal A}$.  If
additionally $\parallel f\parallel =1$, this form is called a
{\em state}.  There exists a correspondence between (equivalence
classes) of representations of ${\cal A}$ in a Hilbert space and
states on $ {\cal A}$.  If the representation $\pi$ is nonzero
and irreducible, the corresponding state is the {\em pure
state\/} (for details see \cite[pp.140-149]{Murphy}).
\par
The algebra ${\cal A}=C_c^{\infty}(G,{\bf C})$ is
noncommutative, and consequently nonlocal in the sense that it
has no maximal ideals.  One can regard it, in analogy with the
commutative case, as describing a certain space, usually called
a {\em noncommutative space}.  We shall also speak of the {\em
associated space\/} with the algebra ${\cal A}$.  Such a space
can be identified with the set $Prim{\cal A}$ of all primitive
ideals of the algebra $ {\cal A}$.  Since this algebra encodes
information on the structure of space-time with singularities we
could say that the space-time with singularities is a
``pointless'' space:  the information on individual points of
space-time has been lost, but the information about the
``structure of singularities'' has been gained.  We could say
that both ``singular'' and ``nonsingular'' states (modelled by
the states on the algebra $ {\cal A}$ in the mathematical sense)
are on equal footing.
\par
In the next section, we shall give a series of theorems
characterizing the existence of singularities of various types
in terms of the noncommutative algebra ${\cal A} =
C^{\infty}_c(G, {\bf C})$.
\par
\section{The Existence of Singularities}
{\bf Lemma 1.} For every $q\in \overline{OM}$, if $
s\in {\cal A}_{proj}$, then 
$$\pi_q(s)(\xi) = k(q) T(\xi )$$
where $k(q)\in {\bf C}$ is the value of the constant function
$s$ on the fiber $\pi_{\overline{OM}}^{-1}(q)$, and $T(\xi ) =
\int_{G_q}\xi (\gamma_1^{-1}\gamma )$.
\par 
{\bf Proof.} Let $s\in {\cal A}_{proj}$. Representation
(\ref{repr}) gives
\[\pi_q(s)(\xi )=s_q*\xi = k T(\xi)(\gamma) \]
for $\xi\in L^2(G_q)$.
$\Box$
\par
The function $s$ from the above proposition is ``constant on a
given fibre'', but need not be constant on different fibres
unless a malicious singularity is present.
\par
{\bf Theorem 1.}  ${\cal A}_{proj}\simeq 
{\bf C}$ if and only if the space-time associated with 
the algebra ${\cal A}$ contains at least one malicious
singularity. In such a case  
$$\pi_q(s)(\xi) = k T(\xi )$$
where $k\in {\bf C}$ is a value of a constant function
$s\in \calA_{proj}$ on any fibre
$\pi_{\overline{OM}}^{-1}(q),\; q \in \overline{OM}$,  and $T(\xi ) =
\int_{G_q}\xi (\gamma_1^{-1}\gamma )$.
\par 
{\bf Proof.} The first part of the theorem follows from the
previous analysis in terms of functional algebras and the fact
that $ {\cal A}_{proj}$ is isomorphic with the algebra of
complex valued functions on $
\bar {M}.$ If we replace ${\cal A}_{proj}$ with ${\cal A}_{pro
j}\cap {\cal A}_H$ we should replace {\bf C} with {\bf R} (for
details see \cite{HS94}.
\par
The proof of the second part of the theorem is the same as that
of lemma 1; one should only notice that since
$\calA_{proj}\simeq {\bf C}$, $k$ is constant on all
fibres.$\Box $
\par
The above results can be rephrased in terms of the {\em total\/}
representation of the algebra ${\cal A}$ on a Hilbert space $
L^2(G):=\bigoplus_{q\in \overline{OM}}(G_q)$.  This 
representation 
\[\pi :\,{\cal A}\rightarrow\bigoplus_{p\in \overline{OM}}
\pi_q(s)\]
is defined by
\[\pi (s)=\bigoplus_{q\in \overline{OM}}\pi_q(s).\]
\par
From lemma 1 it follows that 
$$
\pi(s)(\xi ) = \bigoplus_{q\in
\overline{OM}}k(q)T(\xi),
$$
and the second part of theorem 1 asserts that
$$
\pi(s)(\xi ) = k\bigoplus_{q\in
\overline{OM}}T(\xi).
$$
\par
Let us notice that each $s\in {\cal A}_{proj}$ defines the
function $\tilde {s}:\,\overline{OM}\rightarrow {\bf C}$ by
$\tilde{s} = s(q,e)$ where $e$ is the neutral element of
$\Gamma$; $s$ is then a function which, for a given $ q\in
\overline{OM},$ assumes the value $\tilde {s}(q)$ equal to the
value of the constant function $ s\in {\cal A}_{proj}$ on the
fiber $G_q$.
\par
{\bf Theorem 2.} In the space-time associated with the algebra $
{\cal A}$ there is no 
singularity if and only if $\calA_{proj} \simeq C^{\infty }(M,
{\bf C})$ 
\par
{\bf Proof.} Let us assume that in the space-time associated
with \calA \ there is no singularity, and let $s \in
\calA_{proj}$. The function $\tilde{s}: \overline{OM}
\rightarrow {\bf C}$ is constant on fibres of the frame bundle
$\pi_M: \, \overline{OM} \rightarrow M$, and consequently it
defines the smooth complex valued function $f$ on $M$ by $f(x) =
\tilde{s}(q)$, for $x \in M$, where $q$ is any element of the
fibre $\pi_M^{-1}(x)$. Since $\pi_M(q)=x$ one has
$f(\pi_M(q))=\tilde{s}$, i.e., $f\circ \pi_M = \tilde{s}$. The
smoothness of $f$ follows from the fact that $\pi_M:\,
\overline{OM} \rightarrow M$ is a principal, locally trivial
fibre bundle.  It is also clear that each function of
$\calA_{proj}$ can be obtained by lifting a function $f \in
C^{\infty}(M, {\bf C})$.
\par
Now, let us assume that $\calA_{proj} \simeq
C^{\infty}(M, {\bf C})$. From the construction of the algebra
\calA \ one has that $\calA_{proj} \simeq
C^{\infty}(\bar{M}, {\bf C})$. From this fact and from the
above assumption it follows that
$
C^{\infty}(\bar{M}, {\bf C}) \simeq C^{\infty}(M, {\bf C}).
$
Hence $\bar{M} = M$ which means that there are no singularities.
$\Box $
\par
Let ${\cal A}_{\Gamma_0}$ be the family of $\Gamma_
0$-invariant functions belonging to \calA , i. e., the family of 
functions which are constant on the orbits of the action of $
\Gamma_0$.
\par
{\bf Theorem 3.} In the space-time associated with the algebra $
{\cal A}$ there is an 
elementary quasiregular singularity (but there are no stronger 
singularities) if and only if there exists a discrete group $
\Gamma_0$ of isometries of $M$ such 
that ${\cal A}_{proj} \simeq C^{\infty }(M)_{\Gamma_0}$.
\par
{\bf Proof.} Elementary quasiregular singularities are those
which are produced in the procedure of making the quotient of
space-time by a finite subgroup $\Gamma_0$ of its isometries
\cite{EllisSch}. Only $\Gamma_0$-invariant functions pass
through such singularities \cite{HS95AC,Odrzy}.
$\Gamma_0$-invariant functions lift to all functions of ${\cal
A}_{proj}$. $\Box$
\par
Let us notice that, for malicious singularities, ${\cal A}_{proj}$
consists only of constant functions; for elementary quasiregular
singularities ${\cal A}_{ proj}$ consists of
$\Gamma_0$-invariant functions; if there are no singularities, $
{\cal A}_{proj}$ consists of functions isomorphic with all
smooth functions on space-time.  If in the given space-time
there are singularities of various kinds, the strongest
singularity determines the structure of the algebra
\calA . In agreement with the non-local
character of the noncommutative algebra \calA , the above
theorems convey the information about the structure of singular
space-times (space-times with singularities) rather than about
the structure of singularities themselves.
\par
{\bf Theorem 4.}  In the space-time associated with the
algebra $ {\cal A}$ there is a regular singularity (but there
are no stronger singularities) if and only if the groupoid
$G=\overline{OM}\times\Gamma$ is a subspace of a ``larger''
groupoid $\bar {G}=\bar{E}\times\Gamma$, where $\overline{OM}$
is a subspace of constant dimension (in the sense of Sikorski)
of the space $\bar {E}$.  In such a case $ {\cal A}_{proj}$ is a
localization of $\bar {{\cal A}}_{proj}$ where $\bar {{\cal
A}}_{ proj}$ is the subalgebra of projectible functions on $\bar
{G}$, i.e.,  ${\cal A}_{proj}=(\bar {{\cal A}}_{pro j})_G$.
\par
{\bf Proof.}  We remind that a differential space $ (M,$C) is of
constant dimension (in the sense of Sikorski) if (1) dim$
T_xM=n$ for every $x\in M;$ (2) the module of vector fields
${\cal X} (M)$ is locally free of rank $n$.  In agreement with
the definition of regular singularities such singularities
originate if we cut off a space-time $M$ from the larger
space-time $\tilde{ M}$ so as not to alter its dimension (for
the construction see \cite{HS91}).  This implies that also
$\overline{OM}$$\times\Gamma$ is a subspace of constant
dimension of $\bar{E}\times\Gamma$.
\par
The second part of the theorem is a consequence of the following
implication
$$(C^{\infty}(M)=C^{\infty}(\tilde {M})_M)\Rightarrow 
({\cal A}_{proj}=(\tilde {{\cal A}}_{proj})_G)$$
where $C^{\infty }(\tilde {M})_M$ denotes the localization of
$C^{\infty }(\tilde {M})$ to $M$ (see, beginning of Section 2)
$\Box $ 
\par
It can be seen from the above that regular singularities are
very mild (they can hardly be called singularities):  they do
not change the family ${\cal A}_{proj}$ but only narrow its
domain.
\par
\section{Perspectives} 
Although the analysis carried out in the present paper dealt
with classical singularities, i. e., without taking into account
quantum gravity effects, after making this analysis the question
immediately arises: How would quantum gravity phenomena affect
the obtained results? Usually, one considers two, mutually
excluding themselves, possibilities: either the theory of
quantum gravity, when finally discovered, will remove
singularities from our picture of the universe, or singularities
will remain unaffected by this theory. The first possibility
has, in the last years, become a sort of common wisdom. It
seems, however, that the results of the present paper open
another way of looking at this problem.
\par
As we have seen, the structure of space-time with singularities
is encoded in the noncommutative algebra ${\cal A}
=C_c^{\infty}(G,{\bf C})$.  This algebra is nonlocal, in the
sense that it contains no information about points and their
neighborhoods.  Consequently, singularities cannot be regarded
as points in space-time (or in some other space).  This
conclusion remains in agreement with the standard understanding
of singularities which are usually treated not as points of
space-time, but rather as its ideal points or points of its
singular boundary.  As we have seen, one can meaningfully speak
of (pure) states of the algebra $ {\cal A}$, but there is no
distinction between its singular and non-singular states. This
corresponds to the fact that the physical system (the very early
universe), modelled by the algebra $ {\cal A}$, can occupy
various states, none of which is more singular than the others.
In other words, in this mathematical context, the question on
the existence or non-existence of singularities does not arise.
\par
We could speculate that this mathematical formalism is not only
an artificial tool to deal with classical singularities, but it
also somehow reflects physical regularities of the quantum
gravity era.  The fact that the states on the algebra ${\cal A}$
are represented as algebras of bounded operators in a Hilbert
space (which is typically a quantum structure) can be viewed as
a hint that the algebra $ {\cal A}$ is indeed somehow related to
quantum phenomena.  In fact, there are several attempts to
create a quantum gravity theory based on noncommutative geometry
(see, for instance, \cite{Chamsed,Con96,Hajac,HSL,Sitarz}), but
the above proposal is independent of the particulars of any of
them.
\par
We shall only assume that the algebra ${\cal A}
=C_c^{\infty}(G,{\bf C})$ contains some information about the
pre-Planck era of the universe, and we shall explore some
possibilities hidden in this assumption.  In spite of the fact
that geometry determined by ${\cal A}$ is nonlocal, one can
meaningfully speak of certain physical properties of the system
modelled by it.  As we have seen in Section 3,  the algebra
${\cal A}$ can be completed to the $C^{*}$-algebra. This is
important since $C^{*}$-algebras, in the noncommutative setting,
generalize the standard concept of topology, and the
generalization is so powerful that even non-Hausdorff cases can
be dealt with by using this method (see
\cite[p.79]{Connes}). This could provide a mathematical basis
for some speculations about a ``topological foam'' supposedly
reigning in the quantum gravity regime (see, for instance,
\cite{Isham}). 
\par
Moreover, with every $C^{*}$-algebra ${\cal B}$, represented as
an algebra of operators on a Hilbert space, another algebra,
called a {\em von Neumann algebra}, can be associated; it is
defined as the commutant of its commutant, i.  e.,  $({\cal
B}')'={\cal B}$ where $ {\cal B}'$ is the commutant of ${\cal
B}$.  In our case, such a von Neumann algebra can be obtained as
$ {\cal R}:=(\pi_q({\cal A})')'.$ Given this algebra one can
define the one-parameter group of automorphisms of ${\cal R}$,
$\alpha^{\omega}_ t\in {\rm A}{\rm u}{\rm t}{\cal R}$ for every
$ t\in {\bf R}$, called {\em modular group}, which depends on
the state $\omega$ on $ {\cal R}$ modulo inner automorphisms
(see \cite{ConRov}).
This means that von Neumann algebras are ``dynamical objects''
in the sense that they encode some (abstract) dynamical
properties of a physical system modelled by a given
noncommutative algebra. This is a remarkable circumstance. In
spite of the fact that the algebra ${\cal A}$ is nonlocal (and
consequently in the physical system modelled by it there is no
space and no time in the usual meaning of these terms), one can
define a noncommutative counterpart of dynamics with the modular
group $\alpha^{\omega}_t$ playing the role of ``generalized
time'' \cite{ConRov,Time}.
\par 
The transition from the noncommutative geometry to the usual
space-time geometry can be thought of as a kind of ``phase
transition'' which happens when the universe passes through the
Planck threshold. Mathematically, this corresponds to the
transition from the noncommutative algebra ${\cal A}$ to its
subalgebra $ {\cal A}_{proj}$ as visualized in the following
schema
\[({\cal A}=C_c^{\infty}(G,{\bf C}))\Rightarrow 
({\cal A}_{proj}\simeq\bar {C}^{\infty}(\bar {
M})).\]
As we have seen in Section 3, ${\cal A}_{proj}$ is isomorphic
with the algebra of smooth (in the sense of differential space
theory) functions on the space-time M with its singular boundary
$\partial M,\;\bar {M}=M\cup\partial M$. In this way, after
passing through the Planck threshold both space-time (in the
usual sense) and classical singularities emerge. We are entitled
to say that classical singularities are produced in the process
of the formation of macroscopic physics. Of course, the same can
be said about final singularities, for instance in the closed
Friedman world model or in the gravitational collapse of a
massive object. On the fundamental level, beyond the Planck
scale, there is no distinction between singular and nonsingular
states. Only from the point of view of the macroscopic observer
one can say that the universe had an initial singularity in its
finite past, and possibly will have a final singularity in its
finite future.
\par

\end{document}